\documentclass[twocolumn,english,superscript address]{revtex4-1}
\usepackage[T1]{fontenc}
\usepackage[latin9]{inputenc}
\usepackage{geometry}
\usepackage{units}
\usepackage{amsmath}
\usepackage{amssymb}
\usepackage{graphicx}

\makeatletter
\usepackage{bbm}
\usepackage{latexsym}
\usepackage{dcolumn}
\usepackage{amsmath}
\usepackage{epsf}
\usepackage{float}
\usepackage{hyperref}


\usepackage{babel}
\begin{document}
\title{Generation, Characterization and Manipulation of Quantum Correlations
in Electron Beams }
\author{Shahaf Asban}
\email{Shahaf.Asban@icfo.eu}

\affiliation{ICFO - Institut de Ciencies Fotoniques, The Barcelona Institute of
Science and Technology, 08860 Castelldefels (Barcelona), Spain}
\author{Javier García de Abajo}
\email{Javier.GarciaDeAbajo@icfo.eu}

\affiliation{ICFO - Institut de Ciencies Fotoniques, The Barcelona Institute of
Science and Technology, 08860 Castelldefels (Barcelona), Spain}
\affiliation{ICREA - Institució Catalana de Recerca i Estudis Avançats, Passeig
Lluís Companys 23, 08010 Barcelona, Spain}
\begin{abstract}
Entanglement engineering plays a central role in quantum-enhanced
technologies, with potential physical platforms that outperform their
classical counterparts. However, free electrons remain largely unexplored
despite their great capacity to encode and manipulate quantum information,
due in part the lack of a suitable theoretical framework. Here we
link theoretical concepts from quantum information to available free-electron
sources. Specifically, we consider the interactions among electrons
propagating near the surface of a polariton-supporting medium, and
study the entanglement induced by pair-wise coupling. These correlations
depend on controlled interaction interval and the initial electron
bandwidth. We show that long interaction times of broadband electrons
extend their temporal coherence. This in turn is revealed through
a widened Hong-Ou-Mandel peak, and associated with an increased entanglement
entropy. We then introduce a discrete basis of electronic temporal-modes,
and discriminate between them via coincidence detection with a shaped
probe. This paves the way for ultrafast quantum information transfer
by means of free electrons, rendering the large alphabet that they
span in the time domain accessible.
\end{abstract}
\maketitle

\section*{introduction}

Quantum degrees of freedom occupy a large parameter space compared
with their classical counterparts. This property renders them challenging
for simulation on classical computers. Nonetheless, it also endows
them with a vast information capacity, useful for novel computational
and metrologic paradigms \citep{Helstrom1969,Helstrom1973,nielsen_chuang_2010}.
\emph{Entangled photon pairs} have long been the work-horse of quantum
enhancement demonstrations in the optical arena, with applications
in metrology \citep{Giovannetti2011,Dowling_2008}, imaging \citep{Beskrovnyy2005,Kolobov1999,Brida2010,Rozena2014,Israel2017,Asban2019}
and spectroscopy \citep{Roadmap_mukamel_2020,Dorfman_review_2017,Schlawin_2018,Asban2019}.
A key concept in generation of such useful states, is initiation of
well-monitored interactions between continuous variables. The latter
exhibit rich entanglement spectra and large state-space on which information
can be recorded and accessed \citep{Mandel1985,Law_2000,Law2004,Mair2001,Fickler2013}.
These concepts have not yet been addressed in the well-established
field of free-electron based metrology techniques, such as spectroscopy
and microscopy \citep{Javier_review_2010}. Designing controlled entanglement
of free-electron sources constitutes the main challenge, and this
is precisely what we address here.

Extraordinary electron-beam-shaping capabilities have been recently
demonstrated in electron microscopes combining light ultrafast optics
elements \citep{Vanacore2018,Vanacore2019,Madan2019}. Revolutionary
concepts such as free-electron qubits \citep{reinhardt2019freeelectron}
and cavity-induced quantum control \citep{Kfir_2019,Wang2020,Kfir2020}
are becoming available, pointing towards the emergence of next-generation
quantum light-electron technologies. While photons maintain coherence
over large distances, electrons decohere rapidly due to their strong
environmental coupling. Combined with the control schemes mentioned
above, this suggests that isolated electrons provide valuable \emph{quantum
probes} when selectively exposed to targets of interest. We show that
electrons passing by polariton-supporting media, can experience geometrically
controlled interaction resulting in entanglement. This effect is closely
related to Amperean pairing of electrons discussed in \citep{Lee_2007,Lee_2014,Schlawin_2019},
shown here to induce an entangled EPR state in the long interaction
time limit. 

\begin{figure}[th]
\begin{centering}
\includegraphics[bb=0bp 0bp 255bp 227bp,clip,scale=0.89]{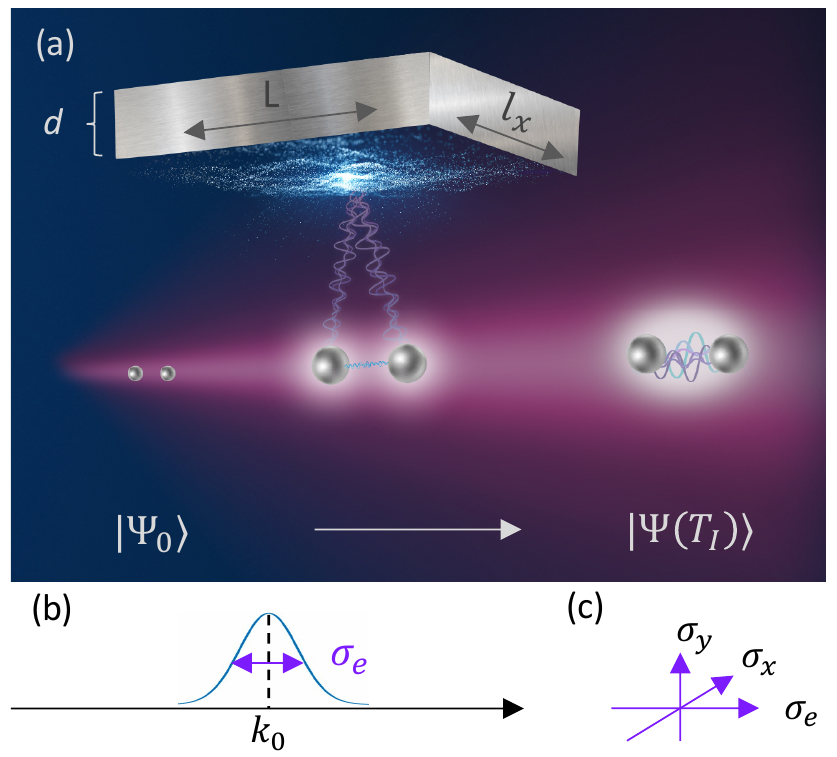}
\par\end{centering}
\caption{\textbf{Physical platform for free-electron pair correlations.} (a)
An uncorrelated electron pair $\vert\Psi_{0}\rangle$ propagates parallel
to the planar surface of a polariton-supporting film of length $L$
along the propagation direction, transverse width $l_{x}\gg L$ and
thickness $d=1\text{ nm}$. (b) Initial distribution of the longitudinal
momentum component, centered around $k_{0}$ with $\sigma_{e}^{2}$
spread. (c) Spatial orientation of the variance in the transverse
momentum spread $\sigma_{x,y}^{2}$. After an interaction time $T_{I}$
, a correlated pair $\vert\Psi\left(T_{I}\right)\rangle$ is obtained.
\label{Fig 1}}
\end{figure}

Here, we study the quantum correlations generated by abrupt interactions
of electron pairs with a neighboring medium, as depicted in Fig. $\text{\ref{Fig 1}}$,
for a controlled time interval $T_{I}$. We explore the transient
state generated by abrupt interactions, as well as the steady state
limit in the pertubative regime. By varying two control parameters
-- interaction time $T_{I}$ and initial electron bandwidth $\sigma_{e}$
-- we effectively scan the degree of entanglement. The entanglement
in the longitudinal dimension is characterized by the Schmidt decomposition
of the wave-function. We then calculate the coincidence probability
and display it versus the degree of entanglement. We denote the resulting
eigenstates \emph{electronic} \emph{temporal modes }(ETMs) in analogy
to their photonic counterparts \citep{Brecht_2015,Raymer_2020}. Finally,
we propose a technique that is useful for real-time discrimination
between ETMs, essential for state tomography and related quantum information
processing applications.

\section*{results}

\subsubsection*{The pair-amplitude}

The electron-pair amplitude is obtained from the underlying electron-polariton
coupling. We consider free electrons traveling with mean momentum
$\boldsymbol{k}_{0},$ as depicted in Fig. $\text{\ref{Fig 1}}$.
The full Hamiltonian is given by three contributions: ${\cal H}={\cal H}_{\text{e}}+{\cal H}_{\phi}+{\cal H}_{\text{e}-\phi}$.
The electrons kinetic term is described by ${\cal H}_{e}$, the electromagnetic
field degrees of freedom combined with the surface polaritons are
contained in ${\cal H}_{\phi}$ \citep{Buhmann_2012,Dung1998} and
the electron-field coupling is ${\cal H}_{\text{e}-\phi}$ (see the
Methods section). Two initially distinguishable electrons illustrated
in Fig. $\text{\ref{Fig 1}}$, are assumed to be prepared in statistically
independent state initially, as reflected by the product state $\vert\Psi_{0}\rangle=\prod_{i=1}^{2}\sum_{\boldsymbol{k}_{i}}\alpha_{s_{i}}^{\left(i\right)}\left(\boldsymbol{k}_{i}\right)\vert\boldsymbol{k}_{i},s_{i}\rangle$.
Here $\vert\boldsymbol{k}_{i},s_{i}\rangle$ represents an electron
state of momentum $\boldsymbol{k}_{i}$ and spin $s_{i}$. The single-electron
amplitude $\alpha_{s_{i}}^{\left(i\right)}\left(\boldsymbol{k}_{i}\right)$
is determined by the preparation process, and we assume it to be a
Gaussian centered around $\boldsymbol{k}_{0}$ along the propagation
axis in our calculations. The opposite spin polarizations allows one
to address each electron separately, and play a central role in quantum
enhanced metrology protocols (elaborated in the Methods section).
As the electrons pass in vicinity to the film, they exchange energy
via the medium. The interaction mediated by the polaritons decays
exponentially with the distance from the medium, validating the use
of pertubative approach (see Sec. S1 of the SI). Expanding the evolution
in the interaction picture to second order, we obtain the electron-pair
wave-function in its generic form 

{\small{}
\begin{equation}
\vert\Psi^{\left(2\right)}\rangle_{\lambda}=\sum_{\boldsymbol{k}_{1}\boldsymbol{k}_{2}}\Phi_{s_{1}s_{2}}^{\lambda}\left(\boldsymbol{k}_{1},\boldsymbol{k}_{2}\right)\vert\boldsymbol{k_{1}},s_{1};\boldsymbol{k_{2}},s_{2}\rangle,\label{General pair amplitude}
\end{equation}
}{\small\par}

\noindent where $\lambda$ labels a set of control parameters. In
the present configuration, $\lambda$ parametrizes the dimensionless
interaction time $T_{I}$ and the initial electron bandwidth $\sigma_{e}$.
We are interested in the dynamics of the longitudinal component of
the electron pair. By tracing the transverse momenta, we obtain an
expression for $\Phi_{s_{1},s_{2}}^{\lambda}\left(k_{1},k_{2}\right)$,
which we denote as the \emph{pair amplitude} (see Eq. $\text{\ref{Pair Amplitude}}$
in the Methods section). The pair amplitude exhibits continuous variable
entanglement throughout most of the explored parameter space. 
\begin{figure*}[t]
\begin{centering}
\includegraphics[bb=0bp 0bp 1160bp 503bp,clip,scale=0.39]{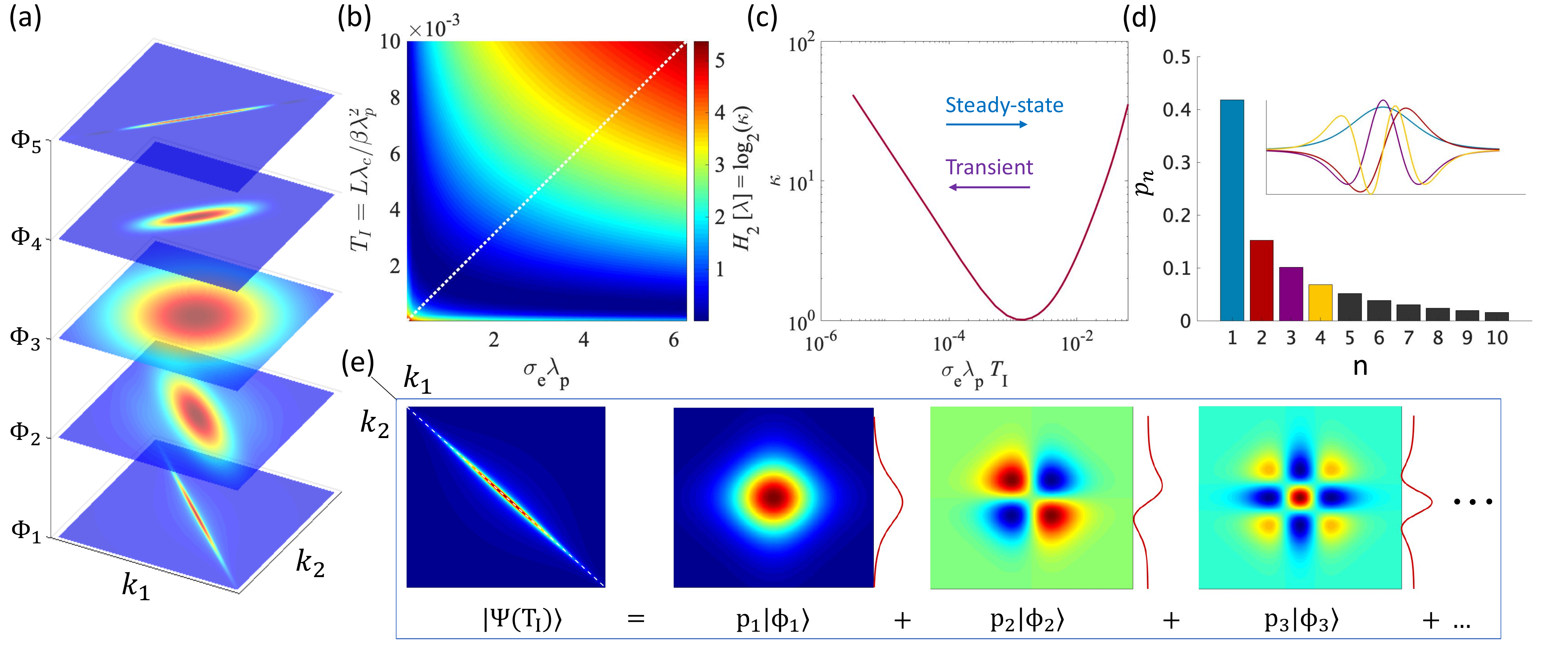}
\par\end{centering}
\caption{\textbf{Entanglement characterization. }(a) The bare pair amplitude
$\Phi_{s_{1},s_{2}}^{\lambda}\left(k_{1},k_{2}\right)$ is presented
for selected control-parameter values, covering the key areas of the
dynamical range. The amplitudes labeled $\Phi_{i}$ are calculated
at the dimensionless interaction times $T_{I}=\left(10^{-2},10^{-3},10^{-5},10^{-5},10^{-5}\right)$,
with bandwidths $\sigma_{e}=\frac{2\pi}{\lambda_{p}}\left(2,2,2,\nicefrac{1}{2},\nicefrac{1}{20}\right)$,
respectively. (b) Collision entropy versus $T_{I}$ and $\sigma_{\text{e}}$,
covering the entire dynamical range from correlated $\left(\Phi_{1}\right)$
to anticorrelated $\left(\Phi_{5}\right)$ momenta. (c) Variation
of the Schmidt number $\kappa$ along the dotted curve displayed in
panel (b), exposing the short time mode-meshing of narrow-band electrons.
(d) Schmidt spectrum of the amplitude displayed in panel (e) $\left(\kappa\approx6\right)$.
The corresponding eigenstates are displayed in the inset with matching
colors. (e) Pair amplitude in joint momentum space, where we display
the first (lowest-order) three modes. \label{Fig2}}
\end{figure*}

\subsubsection*{Entanglement spectrum and ETMs}

It is useful to explore the parameter space of the pair amplitude
by performing a Schmidt decomposition. The Schmidt-Mercer theorem
allows us to express an inseparable state as a superposition of separable
ones,

{\small{}
\begin{equation}
\Phi_{s_{1}s_{2}}^{\lambda}\left(k_{1},k_{2}\right)=\sum_{n}\sqrt{p_{n}}\psi_{n}\left(k_{1}\right)\phi_{n}\left(k_{2}\right),
\end{equation}
}{\small\par}

\noindent where spin labels are omitted for brevity. The longitudinal
eigenstates $\left\{ \psi_{n},\phi_{n}\right\} $ appear in pairs
of ETMs. If the state $\psi_{n}$ is detected, its counterpart occupies
the state $\phi_{n}$ with absolute certainty. The eigenvalues $p_{n}$
reflect the probability of detecting the $\text{n}^{\text{th}}$ mode. 

The joint momentum-representation of the pair amplitude is displayed
in Fig. $\text{\ref{Fig2}}$a for selected values of the control parameters
(i.e., a dimensionless interaction time $T_{I}$ and the electron
bandwidth $\sigma_{\text{e}}\lambda_{p}$). The dimensionless interaction
time is given by $T_{I}=\nicefrac{L\lambda_{C}}{\beta\lambda_{p}^{2}}$,
where $\beta=\nicefrac{v}{c}$ is the electron velocity relative to
of the speed of light, $\lambda_{C}$ is the electron Compton wavelength,
$L$ is the length of the medium along the main propagation direction
and $\lambda_{p}$ is the polariton wavelength in the film (see Sec.
S1 of SI). We employ the collision (Rényi) entropy $H_{2}\left[\lambda\right]=-\log\left(\sum_{n}p_{n}^{2}\right)$
and Schmidt number $\kappa\equiv2^{H_{2}\left[\lambda\right]}$ as
measures for entanglement \citep{Renyi_1961,Law2004,Law_2000,Muller_2013}.
The entropy quantifies the degree of uncertainty with respect to the
instantaneous ETM, while $\kappa$ is the effective number of participating
ETMs. Sweeping the control parameters throughout the entire dynamical
range, reveals two opposite highly correlated regimes, as depicted
in Figs. $\text{\ref{Fig2}}$b-c. For large $\sigma_{\text{e}}\lambda_{p}$
and $T_{I}$ we observe increasing entanglement and correlated momenta
due to energy conservation combined with long exchange times. The
corresponding amplitude is captured in $\Phi_{1}$ of Fig. $\text{\ref{Fig2}}$a,
in agreement with the results reported for photonic counterpart \citep{Law_2000}.
For short interaction times, narrow-band electrons present anticorrelated
momenta. Because of the short interaction time, large energy fluctuations
are introduced in the joint system frame (electrons+film), enabling
a wide range of anticorrelated momenta visible in $\Phi_{5}$ of Fig.
$\text{\ref{Fig2}}$a. In this regime, we observe a rapidly growing
degree of entanglement, captured by growing number of participating
ETMs. The transition between the positively and negatively correlated
regimes, is characterized by a very low Schmidt number $\left(\kappa\approx1\right)$.
This corresponds to an almost separable state, for which a single
ETM is required, corresponding to $\Phi_{3}$ in Fig. $\text{\ref{Fig2}}$a.
In this regime, the electrons can be regarded as approximately disentangled
for all practical purposes. The entanglement spectrum plotted in Fig.
$\text{ \ref{Fig2}}$d is obtained from the Schmidt decomposition
of the amplitude displayed in the extreme left of panel (e), for which
$\kappa\approx6$. The first (lowest-order) three ETMs are visualized
along with their corresponding cross-sections. 

\begin{figure}[th]
\begin{centering}
\includegraphics[bb=0bp 0bp 482bp 434bp,clip,scale=0.46]{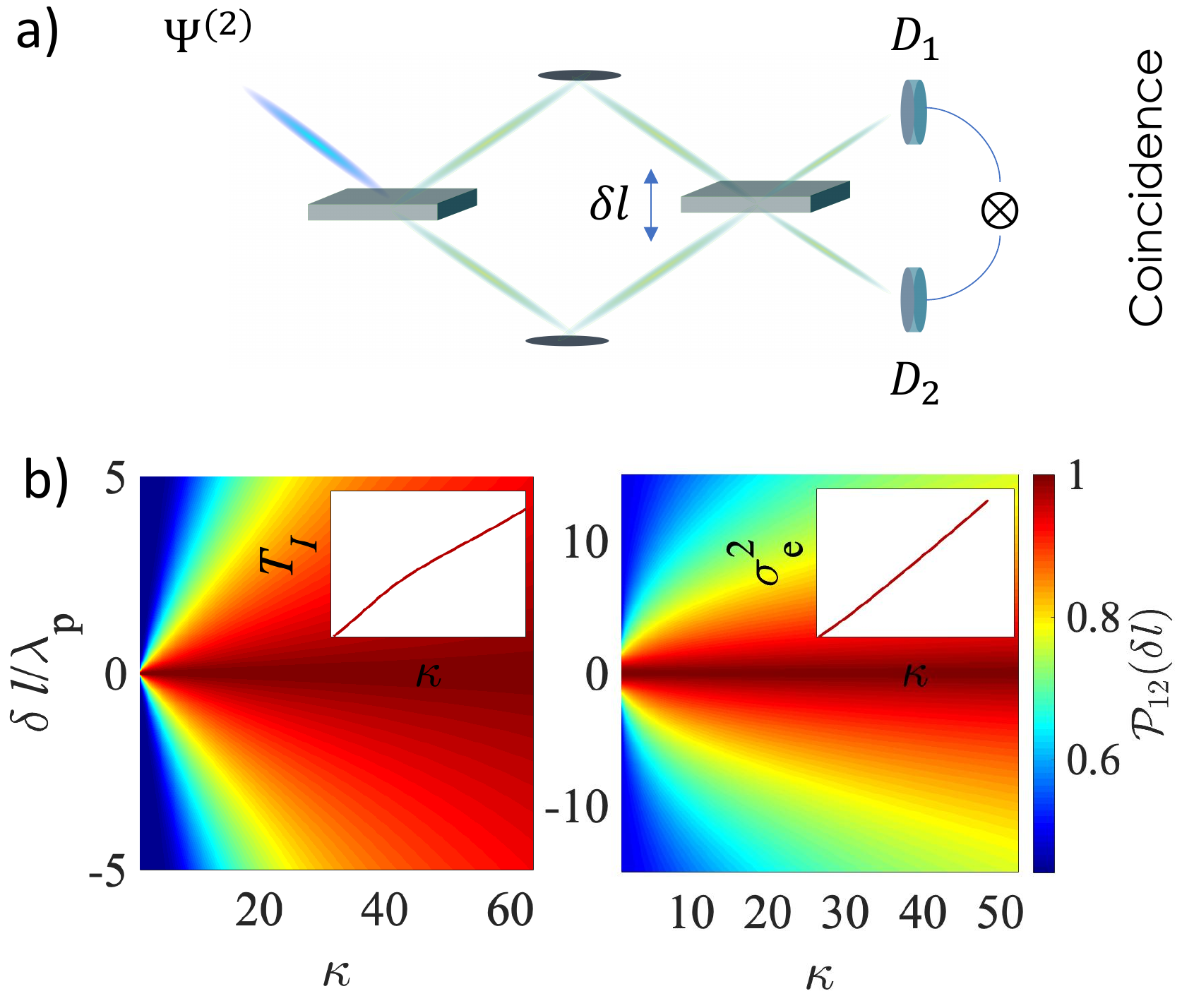}
\par\end{centering}
\caption{\textbf{Coincidence detection.} (a) Incoming electron pairs are separated
by an electron beam splitter (BS), and subsequently combined by another
BS with controllably scanned position, providing a relative path difference
$\delta l$. The two output ports $D_{1}$ and $D_{2}$ are measured
in coincidence. (b) Coincidence probability ${\cal P}_{\boldsymbol{12}}$
for varying path difference $\nicefrac{\delta l}{\lambda_{p}}$ and
Schmidt number (degree of entanglement). The left panel corresponds
to varying interaction time for fixed $\sigma_{e}=\nicefrac{4\pi}{\lambda_{p}}$.
In the right panel the dimensionless time is fixed to $T_{I}=5\times10^{-3}$
while the initial bandwidth is scanned. The insets show the relations
between the control parameters and the Schmidt number. \label{Fig3}}
\end{figure}

\subsubsection*{Coincidence detection}

A common approach to probe quantum correlations is by measuring the
coincidence probability ${\cal P}_{\boldsymbol{12}}\left(\delta l\right)=\int dt\,d\tau\left\langle \Psi_{\boldsymbol{1}}^{\dagger}\left(t\right)\Psi_{\boldsymbol{2}}^{\dagger}\left(t+\tau\right)\Psi_{\boldsymbol{2}}\left(t+\tau\right)\Psi_{\boldsymbol{1}}\left(t\right)\right\rangle $,
assuming an experimental setup as sketched in Fig. $\text{\ref{Fig3}}$a.
We consider balanced beam splitters (BSs) and obtain 

\begin{equation}
{\cal P}_{12}\left(\delta l\right)=\frac{1}{2}+\frac{1}{2}\sum_{nm}\sqrt{p_{n}p_{m}}\left|{\cal I}_{nm}\left(\delta l\right)\right|^{2},\label{Coincidence sum of ETMs}
\end{equation}

\noindent where $n,m$ label ETMs, and ${\cal I}_{nm}\left(\delta l\right)=\int dk\:\phi_{n}^{*}\left(k\right)\phi_{m}\left(k\right)e^{-\frac{i}{\hbar}E_{k}\nicefrac{\delta l}{\boldsymbol{v}}}$
(see Sec. S3 of the SI). Figure $\text{\ref{Fig3}}$b displays ${\cal P}_{12}\left(\delta l\right)$
as a function of the BS displacement, arranged in growing degree of
entanglement. The probability ranges from $\nicefrac{1}{2}$ (completely
random) to unity (utterly antibunched) as expected from fermionic
Hong-Ou-Mandel interference, usually revealed by a Pauli dip in matter
systems \citep{Bocquillon2013,Giovannetti2006,Neder2007}. In the
left panel of Fig. $\text{\ref{Fig3}}$b we scan $T_{I}$ while fixing
$\sigma_{\text{e}}=\nicefrac{4\pi}{\lambda_{p}}$ in the broadband
range. Interestingly, we find that the for higher degree of entanglement
the probability peak extends over a wider range of $\delta l$. This
can be attributed to temporal expansion of the electron wave-function
due to long interaction times. In the inset we see that $\kappa$
grows with $T_{I}$ in a piecewise linear manner, providing a valuable
design tool for a desired target state (see Sec. S3 of the SI). On
the right panel $\sigma_{\text{e}}$ is varied while $T_{I}$ is fixed
in the long-interaction range and a similar behavior is found. We
find that $\kappa\propto\sigma_{\text{e}}^{2}$, which is a direct
consequence of the initial Gaussian wave-packet, together with the
emergent linear relations of $\kappa$ and the interaction time. 

\begin{figure}[th]
\begin{centering}
\includegraphics[scale=0.65]{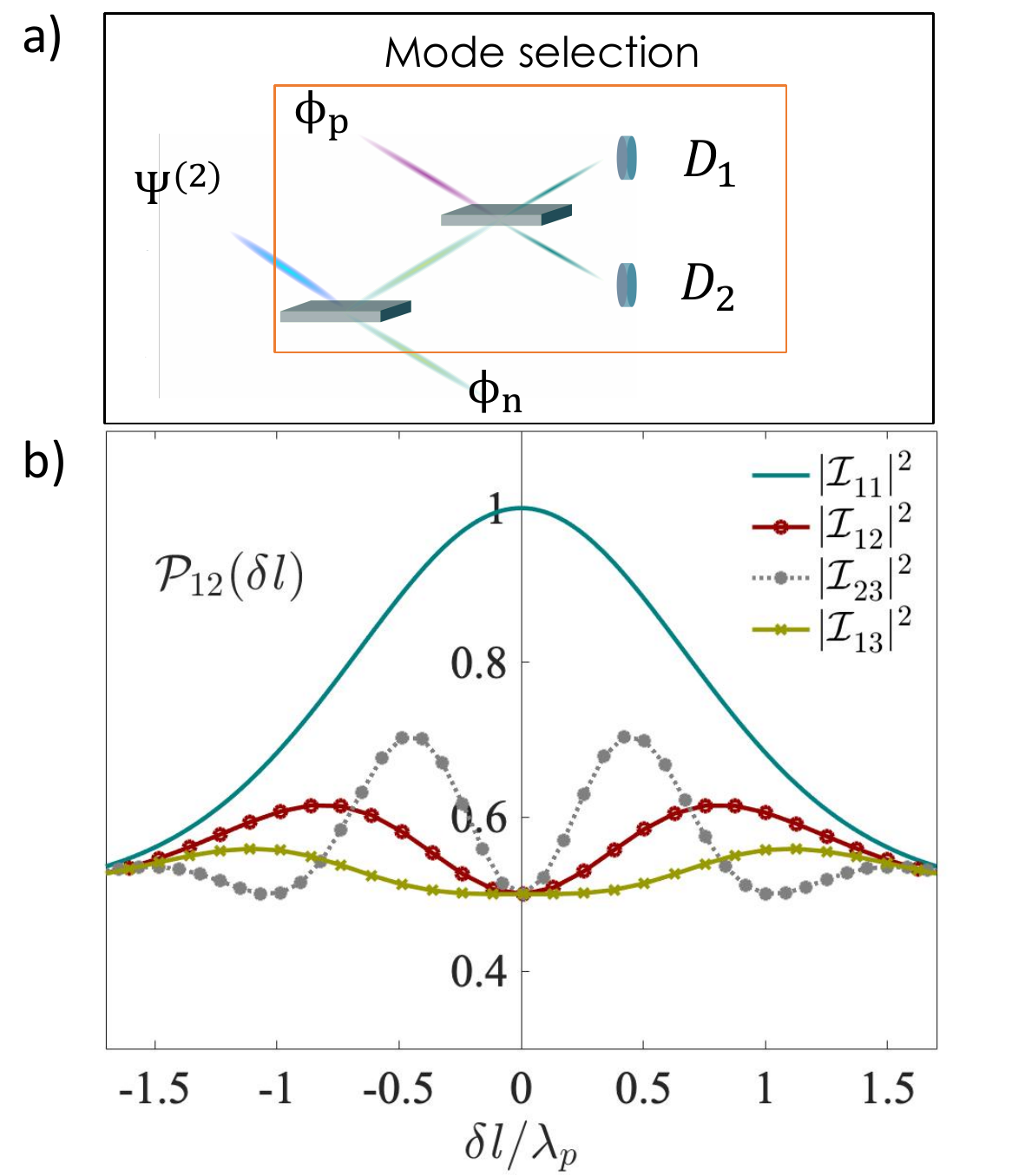}
\par\end{centering}
\caption{\textbf{ETM discrimination.} (a) An incoming electron pair prepared
in a superposition of ETMs is separated by a first BS, then combined
with a (shaped) probe mode $\phi_{p}$ and finally measured in coincidence.
(b) Coincidence outcomes of the probe with three possible incoming
modes $n,p\in\left\{ 1,2,3\right\} $, as a function of path difference
$\delta l$. The interference pattern displays increased response
for identical probe and incoming ETM. \label{Fig4}}
\end{figure}

In Fig. $\text{\ref{Fig4}}$a, the coincidence detection of the \emph{instantaneous}
incoming ETM with a known probe mode labeled $\phi_{p}$ is presented.
The first three ETMs are extracted from the Schmidt decomposition
of the amplitude displayed in Fig. $\text{\ref{Fig2}}$e. These modes
are the eigenstates of the reduced single-electron density matrix,
therefore in each realization one such mode is detected with probability
$p_{n}$. When the incoming mode matches the shaped probe mode, the
coincidence signal exhibits a peak for a vanishing path difference,
as depicted in Fig. $\text{\ref{Fig4}}$b. By counting the appearance
rate of each mode separately we can deduce the probability vector
$p_{n}$, and thus characterize the quantum state. Beyond state tomography
this could also be used in coincidence with parallel operations on
its ETM-twin, realizing more sophisticated information processing
protocols. 

\section*{discussion}

Near-fields evolving at the surface of polariton supporting materials
provide a novel approach to generate and shape quantum correlations
in charged particles, and in particular in free electrons. While such
pairing mechanisms are suppressed in matter due to ambient noise (e.g.,
thermal), electrons structured in a beam undergo significantly less
scattering events, thus enabling coherent interactions to persist
over longer space-time intervals. We have shown that electron pairs
near polariton-supporting material boundaries undergo nontrivial coupling
that generates entanglement. Such correlations are mathematically
expressed by the apparent inseparability of the pair amplitude in
Eq. $\text{\ref{General pair amplitude}, giving rise to the results displayed in Fig. \ensuremath{\text{\ref{Fig2}}}}$.
The Schmidt decomposition allows us to express the pair amplitude
using a set of factorized states, providing useful measures for bipartite
entanglement \citep{Ekert_1995,Parker_2000,Law_2000,Law2004,Giedke_2003,Straupe_2013,Laskowski_2012,Sciara2017,Giddings_2018}.
This framework reveals simple relations between the control parameters
and the resulting evolution of quantum correlations of the above setup.
Such properties are desirable for entanglement engineering. 

The large Hilbert-space dimensionality occupied by the ETMs, renders
them as appealing ultrafast quantum information carriers. This has
potential applications in quantum-enhanced electron metrology, as
proposed using optical setups \citep{Giovannetti_2011,Schlawin_2018}.
For example, measuring the momentum of one of the electrons in the
pair and the position of the other, one may obtain superesolved imaging.
Such class of quantum enhancements benefits from the fact that single-particle
(local) observables are not Fourier conjugates of the (extended) composite
state. This work raises multiple open questions concerning vicarious
temperature effects, the imprint of the medium topology on the entanglement
spectrum, entanglement along the transverse plane and higher order
matter-electron quantum correlations. These are just a few examples
of the emerging field of quantum free-electron metrology. From the
information theoretic point of view, the large alphabet spanned by
ETMs, promotes their candidacy for electron-beam quantum information
processing and communication tasks. This raises questions regarding
information capacity of the channel in the presence of noise, providing
a direction for future study.

\section*{methods}

\subsubsection*{Electron source}

First, it is useful to discuss the importance of distinguishability
in the initial product state $\vert\Psi_{0}\rangle=\prod_{i=1}^{2}\sum_{\boldsymbol{k}_{i}}\alpha_{s_{i}}^{\left(i\right)}\left(\boldsymbol{k}_{i}\right)\vert\boldsymbol{k}_{i},s_{i}\rangle$.
In order to benefit from the entangled state in a quantum-metrological
sense, one crucially relies on the ability to address each of the
particles separately, thus, exposing nonlocal effects (e.g., photon
polarization \citep{Law_2000,Dorfman_review_2017}). By addressing
each particle separately using a specified degree of freedom (here
the spin), one can measure conjugate quantities -- such as the momentum
of one and the position of the other -- with increased sensitivity
\citep{Howell_2004,Saunders_2010}. (Complementary to quantum correlations
of indistinguishable fermions revealed by the Slater rank \citep{Schliemann_2001}.)
We consider the initial state to be prepared using a spin polarized
electron source, allowing separate single particle manipulation prior
to the interaction \citep{Kohashi_2009,KOHASHI_2015,Kuwahara_2011}.

\subsubsection*{Pair-amplitude derivation}

The pair amplitude is obtained pertubatively in the interaction picture
(Sec. S1 of the SI). The full Hamiltonian of the system contains three
contributions: ${\cal H}={\cal H}_{\text{e}}+{\cal H}_{\phi}+{\cal H}_{\text{e}-\phi}$.
The electrons kinetic term is given by ${\cal H}_{e}=\sum_{\boldsymbol{k,s}}\epsilon_{\boldsymbol{k}}c_{\boldsymbol{k,}s}^{\dagger}c_{\boldsymbol{k,}s}$,
where the operator $c_{\boldsymbol{k},s}\left(c_{\boldsymbol{k},s}^{\dagger}\right)$
creates (annihilates) an electronic mode with momentum $\boldsymbol{k}$
and spin $s$ obeying the anticommutation relations $\left\{ c_{\boldsymbol{k},s},c_{\boldsymbol{k}',s'}\right\} =\delta_{\boldsymbol{k}\boldsymbol{k}'}\delta_{ss'}$.
The term ${\cal H}_{\phi}$ describes the electromagnetic-field degrees
of freedom combined with the surface polaritons in the framework of
macroscopic quantum electrodynamics \citep{Buhmann_2012,Dung1998}.
The electron-field coupling is expressed using the Hamiltonian \citep{dressel_2002,Dung1998,Asban_2020}

\begin{equation}
{\cal H}_{\text{e}-\phi}=\frac{e\lambda_{C}}{2\pi}\sum_{\mathbf{k},\mathbf{q},s}c_{\mathbf{k}+\mathbf{q},s}^{\dagger}c_{\mathbf{k},s}\mathbf{k}\cdot\mathbf{A}\left(\mathbf{q}\right),
\end{equation}

\noindent where $\lambda_{C}=\nicefrac{h}{m_{\text{e}}c}$ is the
Compton wavelength of the electron, while $e$ and $m_{\text{e}}$
are its charge and mass, respectively. We have employed the Weyl-gauge,
setting the scalar potential to zero and introduced $\mathbf{A}\left(\mathbf{q}\right)$,
the vector field operator in momentum space. The vector-field in macroscopic
quantum electrodynamics is expressed in terms of the Green tensor,
encapsulating the geometric and spectral properties of the medium.
Proceeding to calculate the first nontrivial order (second), we obtain
the general form of Eq. $\text{\ref{General pair amplitude}}$. We
consider Gaussian initial states of mean distance $y_{0}=5\,\text{nm}$
from the thin film and $\sigma_{y}=0.5\,\text{nm}$ (see Fig. $\text{\ref{Fig 1}}$).
Taking the long $l_{x}$ limit and choosing $\sigma_{x}=\frac{2\pi}{\lambda_{p}}$,
we trace the transverse components and obtain

{\small{}
\begin{align}
\Phi_{s_{1}s_{2}}^{\lambda}\left(k_{1},k_{2}\right) & ={\cal N}^{-\nicefrac{1}{2}}\int dq\,\text{sinc}\left[\frac{\hbar q}{m}\left(k_{1}-k_{2}\right)T\right]\nonumber \\
\times\qquad & \alpha_{s_{1}}^{\left(1\right)}\left(k_{1}-q\right)\chi\left(q\right)\alpha_{s_{2}}^{\left(2\right)}\left(k_{2}+q\right).\label{Pair Amplitude}
\end{align}
}{\small\par}

\noindent Here, ${\cal N}$ is a normalization constant, $T$ is the
interaction time, and $\chi\left(q\right)$ is obtained by tracing
the lateral wave vector $\boldsymbol{q}_{\parallel}=\left(q_{x},q_{y}\right)$
in the interaction picture (Sec. S1 of SI). Additionally, we have
invoked the nonrecoil approximation for small momentum exchanges relative
to $k_{0}$, resulting in a linear electron-energy exchange $\epsilon_{\boldsymbol{k+q}}-\epsilon_{\boldsymbol{k}}\approx\hbar\boldsymbol{q}\cdot\boldsymbol{v}$,
where $\boldsymbol{q}$ is the polariton wave vector and $\boldsymbol{v}$
is the electron velocity.

\subsubsection*{ETMs calculation}

To find the set of ETMs $\left\{ \psi_{n},\phi_{n}\right\} $ and
their weights $p_{n}$, we solve the integral eigenvalue equations
$p_{n}\psi_{n}\left(k\right)=\int dk'\,K_{1}\left(k,k'\right)\psi_{n}\left(k'\right)$
and $p_{n}\phi_{n}\left(k\right)=\int dk'\,K_{2}\left(k,k'\right)\phi_{n}\left(k'\right)$
(Sec. S2 of SI). The kernels, which are found from the reductions
$K_{1}\left(k,k'\right)=\int dk_{2}\,\Phi_{s_{1}s_{2}}^{\lambda}\left(k,k_{2}\right)\Phi_{s_{1}s_{2}}^{\lambda*}\left(k',k_{2}\right)$
and $K_{2}\left(k,k'\right)=\int dk_{1}\,\Phi_{s_{1}s_{2}}^{\lambda}\left(k_{1},k\right)\Phi_{s_{1}s_{2}}^{\lambda*}\left(k_{1},k'\right)$,
can be interpreted as single-electron correlation functions. To obtain
the Schmidt spectrum and characterize the degree of entanglement,
we discretize the kernels and numerically solve the integral eigenvalue
equations. We have used a $800\times800$ discretization of the kernel
and repeated the procedure for each control parameter separately.
The pair amplitude used for the generation of the kernels involves
integration over the polariton degrees of freedom. We have done this
numerically for each set of control parameters $\lambda$ using straightforward
numerical integration of Eq. $\text{\ref{Pair Amplitude}}$ on a uniform
grid. The step size was varied to satisfy the convergence of the Schmidt
number. The convergence criterion adopted in this scheme is $\text{max\ensuremath{\left\{  2\left(\kappa_{N+1}-\kappa_{N}\right)/\left(\kappa_{N+1}+\kappa_{N}\right)\right\} } \ensuremath{\leq0.05}}$,
where $N$ is the number of data points within a constant range in
the given kernel size. 

\section*{data availability }

The main results of this manuscript are composed of analytical and
numerical calculations. All data generated, analyzed or required to
reproduce the results of this study are included in this article and
its Supplementary Information file.

\bibliographystyle{unsrt}
\bibliography{EEP_bib}

\begin{acknowledgments}
We gratefully acknowledge help from Noa Asban on the graphical illustrations.
This work has been supported in part by the Spanish MINECO (MAT2017-88492-R
and SEV2015-0522), the European Research Council (Advanced Grant 789104-
eNANO), the European Commission (Graphene Flagship 696656), the Catalan
CERCA Program, and Fundació Privada Cellex. 
\end{acknowledgments}

\end{document}